\documentclass{elsart}

\usepackage{graphicx}
\usepackage{bm} 

\begin{document}

\begin{frontmatter}
\title{\bf The balance functions in azimuthal angle is a measure of the transverse
  flow\thanksref{grant}}
\thanks[grant]{Research supported in part by the Polish State
Committee for Scientific Research, grant 2 P03B 059 25}  
\author{Piotr Bo\.zek}

\address{The H. Niewodnicza\'nski Institute of Nuclear Physics, \\
         Polish Academy of Sciences, PL-31342 Krak\'ow, Poland}

\begin{abstract} 
The  charge or barion number balance function in the relative azimuthal angle of
a pair of particles emitted in ultrarelativistic heavy ion collisions
 is studied.
The $\pi^+\pi^-$ and $p\bar{p}$ balance functions are computed using 
thermal models with two different set of parameters, corresponding to a large
freeze-out  temperature  and a moderate transverse flow or a small temperature
and a large transverse flow. The single particle spectra including pions 
from resonance decays are similar for the two scenarios, on the other
hand the azimuthal balance function is very different and could serve as an
independent measure of the transverse flow at the freeze-out.
\end{abstract}


\end{frontmatter}

\vspace{-7mm} PACS: 25.75.-q, 24.85.+p

 The  
balance function in rapidity has been proposed 
as a 
measure of correlations between particles of  opposite charges
\cite{baldef,Bass,jeon}. This includes the electric charge, the barion number
or the strangeness, which we call below charge  correlations, unless
specified explicitely. Elementary processes occurring in heavy ion collisions
conserve the charges which means that   opposite charged particles
 are  produced in pairs. 
One expects that due to such a microscopic local constraint on the production
process of a pair of
particles some correlations in the momenta of the particles appear.

We propose a new observable 
sensitive to the freeze-out characteristics,
 {\it the charge balance function in the 
relative 
azimuthal angle}.
In the following we call this quantity the $\phi-$balance function, 
it is defined 
similarly to the charge balance function in rapidity
\begin{eqnarray}
B^\phi(\delta\phi,\phi) &=& {1\over 2} 
\left\{
{\langle N_{+-}(\delta \phi,\phi) \rangle - \langle N_{++}(\delta\phi,\phi)
 \rangle \over
\langle N_+ (\phi)\rangle} \right. \nonumber \\
& & \left.
+
{\langle N_{-+}(\delta\phi,\phi) \rangle - \langle N_{--}(\delta\phi,\phi) \rangle \over
\langle N_- (\phi)\rangle} \right\},
\label{defphi}
\end{eqnarray}
where, e.g.
 the
 quantity $N_{+-}(\delta \phi,\phi)$ is  defined as the number of pairs
 of particles
with the particle ($+$) flying at an angle $\phi$ and the particle (${-}$)
  at an angle 
$\phi+\delta\phi$. The dependence on the angle $\phi$ (measured with respect 
to the reaction plane)
 could be useful to study the difference between the correlations 
in and out of the reaction plane. 
 Other quantities appearing in Eq. (\ref{defphi}) are defined in an
analogous way.
In the following we study the azimuthally averaged
$\phi-$balance function  
\begin{equation}
B^\phi(\delta\phi)=\int_0^{2\pi} d\phi B^\phi(\delta\phi,\phi) 
\end{equation}
and azimuthally symmetric freeze-out conditions.
Obviously $B^\phi$ has the same normalization condition as the
 balance function in rapidity
\begin{equation}
\int_{-\pi}^{\pi}d\delta\phi B^\phi(\delta\phi) =1 ,
\end{equation}
with the acceptance window and detector efficiency effects modifying 
the above relation
 \cite{phdmsu,jeon}.
The $\phi-$balance function can be studied in ultrarelativistic heavy 
ion experiments for different centrality and transverse momentum cuts.


The STAR Collaboration has presented results on the charge balance function
for Au-Au collisions  at RHIC, indicating a narrow correlation in rapidity
\cite{STARbal,phdmsu},
significantly smaller than observed in elementary particle collisions.
It means that the charges in   heavy ion collisions are produced in a different
way than in elementary processes. In particular, this observation indicates
the creation of charged particle pairs in the late stage of the collision. 

Theoretical estimates of the balance function in rapidity for pion pairs have
been obtained in the framework of thermal models \cite{my1,msu}.
The basic assumption behind those calculations is that the pair formation
occurs in a thermalized local source. It is consistent with a late
hadronization scenario, as the $\pi^+\pi^-$ pair must be created late in
a system which undergoes a strong longitudinal flow \cite{Bass}.
The $\pi^+ \pi^-$ balance
function has two contributions, corresponding to two different
mechanisms for the creation of a pair of  opposite charges  \cite{my1}.
 The first one (the resonance contribution)
 is determined by the decays of neutral
hadronic resonances with a $\pi^+\pi^-$ pair in the final state.
 The second one  (the nonresonant contribution)
is given by the emission of pair of opposite charged particles from a local
thermal source. 
The momenta of the particles in the source rest frame
are given by the thermal distribution, and are  boosted by the velocity
corresponding to the collective flow of the source.
For the calculation of the 
 $p\bar{p}$ correlation we take only the second mechanism, the emission
 from a thermal source.

We compare two different thermal models,
 the single freeze-out model \cite{thermal}
and the boost invariant blast-wave parameterization \cite{starbw}.
 In the first case the kinetic freeze-out happens at the same time as the chemical one. 
The temperature of the latter is fixed by the observed particle ratios at a 
relatively high value of $T_f=165$MeV. The elements of the hypersurface
of emission
 are moving with 
 a collective flow velocity given by the three dimensional Hubble flow. In the transverse 
direction the flow has
 the scaling  form $\beta_r=\frac{r}{t}$ as
 a function of the radial distance $r$,
 with $t=\sqrt{\tau^2+r^2}$, $\tau=7.6$fm, and 
$0<r<\rho=6.7$fm. The parameters $\rho$ and $\tau$ define the size of 
the source and the 
amount of the transverse flow. The average transverse velocity is 
$\langle\beta\rangle=0.5$. 
On the other hand the blast wave model assumes a late kinetic freeze-out, happening some 
time after the chemical processes have ceased. The typical
temperature at the freeze-out 
$T_f=90$MeV
and the parameters of the transverse flow are fixed by a fit to the
transverse momentum spectra of pions, protons and kaons
\cite{starbw}.  The transverse flow in the 
blast-wave parameterization is given by $\beta_r=\beta\left(\frac{r}{r_{max}}
\right)^\alpha$
where the parameters $\alpha=0.82$ and $\beta_r=0.84$ are adjusted to describe the
 spectra in central events at $\sqrt{s}=200$GeV. The freeze-out temperature
 and the flow profile 
depend  on the energy and the centrality of the collision. 
In the following  we take only the quoted above  parameters of the blast-wave model corresponding to central events at the highest 
RHIC energy as representative for thermal models with a late kinetic freeze-out. 

\begin{figure}[tb]
\begin{center}
\includegraphics[width=10cm]{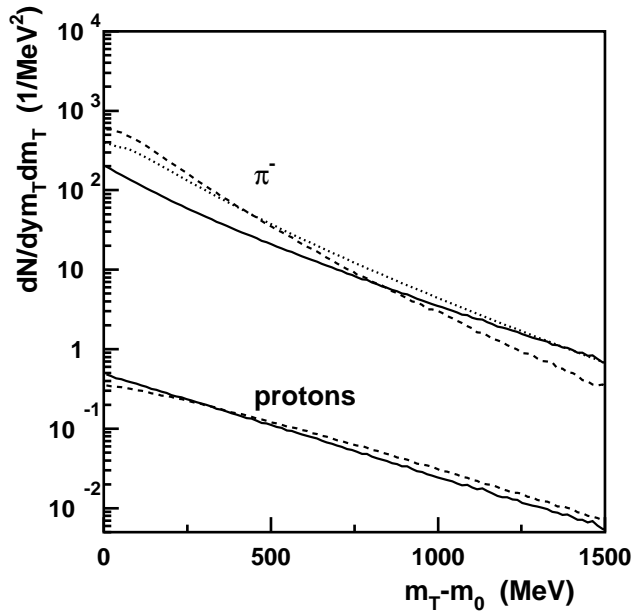}
\end{center}
\caption{Spectra in the transverse mass calculated in  thermal models for two
  different freeze-out conditions. $T_f=90$MeV and $\langle\beta\rangle=0.6$
  (dashed line) and $T_f=165$MeV and $\langle\beta\rangle=0.5$ (solid line)
for pions (upper curves) and protons (lower curves). 
The dotted line denotes the pions obtained at the higher freeze-out
  temperature including pions from the decay of
 resonances. The normalization is arbitrary.} 
\label{trmass}
\end{figure}

Both the single freeze-out model and the blast-wave model fit well
the single -particle transverse momentum spectra. The reason is that in the 
first model a large freeze-out temperature is supplemented with a moderate 
flow $\langle \beta_r\rangle=0.5$ and 
in the second model 
the emission at a small temperature happens in the presence of a significant 
transverse flow $\langle\beta_r\rangle=0.6$.
The resulting spectra of pions and protons are very similar 
(Fig. \ref{trmass}).
Some difference is visible for low momentum pions but in the scenario with a high 
freeze-out temperature, most of the pions come from the decay of resonances.
This effect modifies noticeably the spectra at small momenta. 
The resulting transverse mass spectra for the single freeze-out model 
with resonance decays included is very similar to the blast wave results, 
although the temperatures are very different.
Due to the combined effect of the transverse flow and of the 
 thermal motion with
 the addition of resonance decays it is not possible to determine  the 
freeze-out temperature and the flow from the spectra of light hadrons only.
This lack of sensitivity is at the origin of two different parameterization of the 
freeze-out surface which can describe acceptably the experimental spectra of particles 
produced in ultrarelativistic heavy ion collisions. A simultaneous description
of HBT radii could give an additional constraint \cite{hbt1,hbt2,hbt3}
 but  within both models the 
gross features of the observed correlation radii can be reproduced.
From general principles different space-time emission function can given
similar HBT correlations for on-shell pions.

In thermal models hadron production occurs at the chemical 
freeze-out.
In the  single freeze-out model \cite{thermal} the kinetic and chemical 
freeze-outs happen at 
the same time and the distribution of momenta of a correlated pair of particles
is not significantly changed after its creation.
 If the kinetic freeze-out is delayed, as
in the blast-wave models, one has
 to assume that the two correlated particles are
created in a local thermal ensemble at the chemical freeze-out and
 follow the same thermal 
history until the kinetic freeze-out. By rescattering the two particles 
stay in contact with the same local thermal system, as a consequence the particle momenta 
evolve with the 
decreasing temperature and increasing transverse flow of the system \cite{msu}.
It has been noticed that the charge balance function in rapidity is 
somewhat sensitive to the characteristics of the final state \cite{Bass,STARbal,msu}.
For massive particles the width of the balance function decreases with
 increasing ratio
of the mass to the temperature.
The charge balance
function in rapidity is narrower for pions emitted from a source with a large
transverse velocity or for pions originating from the decay of a fast
resonance \cite{my1,msu}. The  variation 
 of the amount of  transverse flow with
the centrality of the collision can reproduce \cite{msu} the experimentally 
observed narrowing of the balance functions for central events
\cite{STARbal}.
The $\phi-$balance function  measures the charge corelation 
in the relative angle of 
the emitted pair and not the rapidity.
 We show that this quantity is {\it very sensitive}
 to the characteristics of the 
freeze-out and could also give some insight into the 
microscopic process of the charge 
creation.

\begin{figure}[tb]
\begin{center}
\includegraphics[width=10cm]{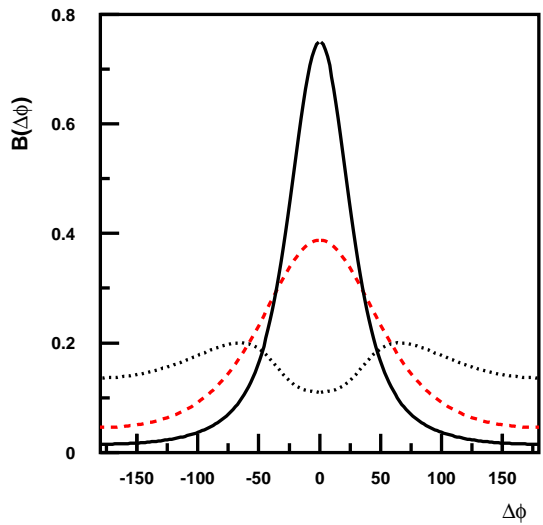}
\end{center}
\caption{Balance functions for  pions in  thermal models
calculated for two different freeze-out conditions~: $T_f=165$MeV,
$\langle\beta\rangle=0.5$ (dashed line for nonresonant
 pions and dotted line for
pions from the decay of a $\rho_0$ resonance)  and for nonresonant pions at 
$T_f=90$MeV,
$\langle\beta\rangle=0.6$ (solid line).}
\label{rhophi}
\end{figure}

The calculation proceeds in a similar manner as for the charge balance function 
in rapidity \cite{my1}. For the $\pi^+\pi^-$ pairs we take explicitely into account the 
pion pairs from the decay of the neutral resonances
\begin{equation}
K_S, \; \rho^0, \; \sigma, \; {\rm and} \; f_0,
\label{res}
\end{equation}
as well pions emitted from the local thermal source.
In the rest frame of the resonance,
pions from a two-body decay are correlated back to back in the azimuthal
angle.
Nonresonant pion are emitted isotropically in the local frame of the source, 
therfore their correlation in the relative  angle is flat. 
Isotropic emission is also a good approximation for the decays with many-body
final states, therfore for the production of   the resonant pion pairs 
we  take  into account only neutral
resonances with two-body decays (\ref{res}).  
 The azimuthal angle 
correlations are modified by the collective flow of the thermal source or by the 
fact that the resonances decay in flight. In Fig. \ref{rhophi} is presented 
the $\phi-$balance function for pions from $\rho$ decays (at $T_F=165$MeV)
 and for nonresonant pions thermally emitted 
with the two freeze-out conditions that we consider. The contribution of 
pions from resonance
decays can be neglected at $T_f=90$MeV and is not taken into account. 
Clearly, pions emitted
with a small temperature have smaller relative momenta and after the boost from the 
rapidly moving source frame to the  laboratory frame their momenta are pointing 
approximately in the 
same direction. The situation is different at the higher freeze-out temperature. First,
 thermally emitted pions have a significant thermal motion and second the
 boost to the laboratory frame is done with a smaller velocity. At
 $T_f=165$MeV, 30\% of pions \cite{share} 
come from the decay of resonances (\ref{res}).
Due to the back to back 
emission the angular distribution of resonance decay products 
is even wider than for the nonresonant pions at $T_f=165$MeV.
It was noted that the effect of resonance decays reduces the elliptic flow of
pions \cite{hirano,hbt1}. If restricted to $\pi^+\pi^-$ correlations and to
the  relative angle of the pair   as for the $\phi-$balance function
this effect is even stronger.

\begin{figure}[tb]
\begin{center}
\includegraphics[width=10cm]{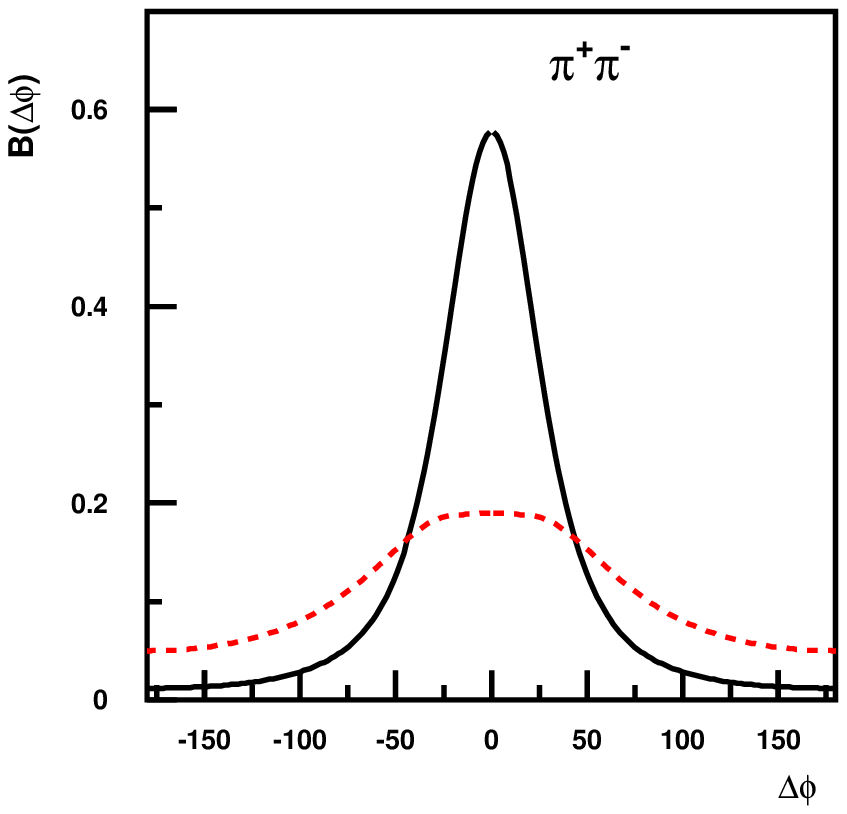}
\end{center}
\caption{Balance functions for pions in     thermal models
calculated for two different freeze-out conditions~: $T_f=165$MeV,
$\langle\beta\rangle=0.5$ (dashed line) and $T_f=90$MeV,
$\langle\beta\rangle=0.6$ (solid line).}
\label{piphi}
\end{figure}

The correlation between the pions is given by a  weigthed \cite{my1}
 sum of the two mechanisms
(nonresonant pions and  decay products of resonances listed in (\ref{res}))
for $T_f=165$MeV and by nonresonant pion pairs for the freeze-out at
$90$MeV. The width of  resulting $\phi-$balance function is very different
(Fig. \ref{piphi}).
At $T_f=165$MeV 
both the nonresonant pion pairs and pion pairs from the decay of resonances
 have a  large angular separation. The contribution of the resonant
 pion pairs makes the $\phi-$balance function show a 
flat correlation between $-50^{\circ}$ and $50^{\circ}$. The width of the
$\phi-$balance function decreases with the mean transverse momentum of the pion
pair.  The average angle  between a $\pi^+$ and a $\pi^-$ is
 $54^{\circ}$ 
for pairs of $1$GeV total transverse momentum 
 and $24^{\circ}$ for the total momentum of $2$GeV if the pions are emitted at
 $T_f=90$MeV, whereas it is $80^{\circ}$ and $50^{\circ}$ respectively 
for pions emitted at the higher
 freeze-out temperature. Of course for very small total momentum of the pair the
 correlation is trivially dominated by back to back emission. 
Changing the mean momentum of the pair the
 contribution of pions from resonance decays changes, being the largest for
 small momenta.

\begin{figure}[tb]
\begin{center}
\includegraphics[width=10cm]{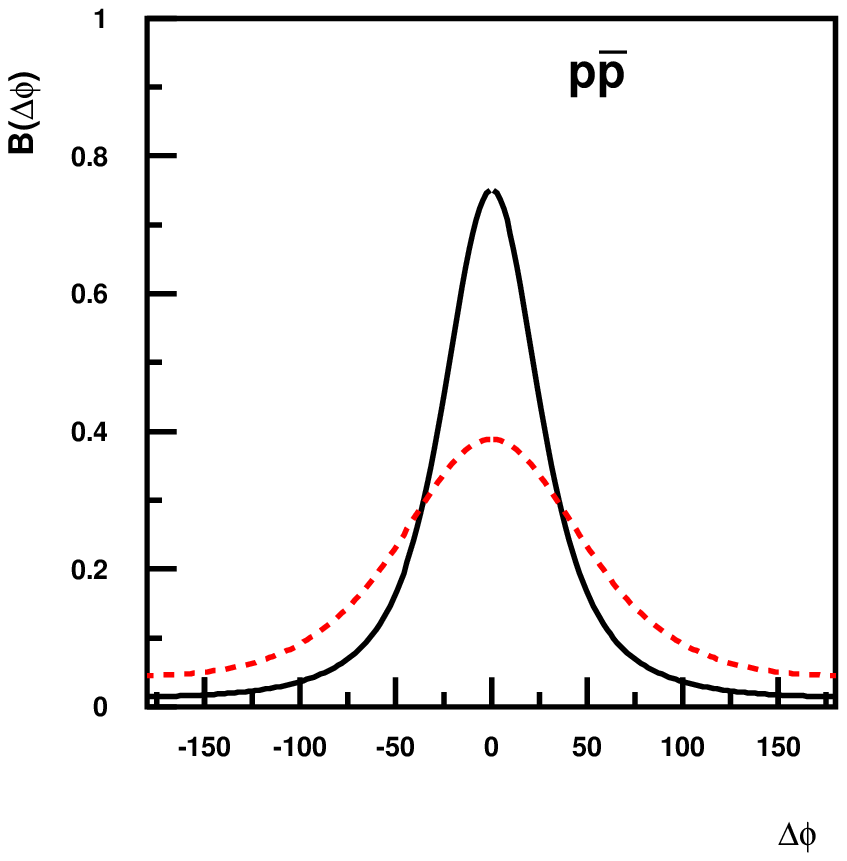}
\end{center}
\caption{Proton-antiproton balance functions 
in  thermal models
calculated for two different freeze-out conditions, as in Fig. \ref{piphi}.}
\label{prophi}
\end{figure}

In Fig. \ref{prophi} the $\phi-$balance function for protons and
antiprotons is shown. 
Very similar behavior appears as for the pion correlations. The emission at
higher temperature and smaller transverse flow produces proton
and antiproton pairs less focused in the azimuthal angle.
Protons which are quite massive 
  are more sensitive to the transverse flow and
have less thermal motion than pions. Also resonance decays are expected 
not to modify significantly  our estimates for the baryonic charge
$\phi-$balance function; fits to the experimental data  
could give direct information on the freeze-out parameters.

In summary, we propose to study   charge-anticharge correlations {\it in the
  azimuthal angle}. The experimental observable
 is defined as the charge balance function in the
 relative  azimuthal angle of the particle pair. 
 The measurement of the $\phi-$balance function in heavy ion collisions could 
serve as another independent constraint on the temperature and the amount of
  the transverse flow at the freeze-out. The effect of an increase of the
  freeze-out temperature and of  a simultaneous  decrease of the transverse
 flow compensate to a large extend in the single particle spectra. 
On the other hand both the increase of the temperature and the reduction of
  the transverse flow make the $\phi-$balance function wider.
Accordingly the width of the $\phi-$balance function could be used independently
 of the particle spectra and HBT radii 
to determine the values of the freeze-out parameters. Explicit estimates of
  the $\phi-$balance function for two thermal models with different freeze-out
temperatures and different transverse flow confirm these expectations both for
$\pi^+\pi^-$ and $p\bar{p}$ pairs. We find that the  most sensitive region
  is for the total momentum of the pair of  around $1$GeV.

Finally we note that the $\phi-$balance function  $B^\phi(\delta \phi,\phi)$
with angular dependence with
respect to the reaction plane (Eq. \ref{defphi}) could give some insight into
 the angular dependence of the transverse flow. The elliptic flow gives a combined
information on the spatial and momentum asymmetries \cite{hbt1,hbt3}. The 
measurement of the angular dependence of the width of 
the $\phi-$balance function
could serve as a probe  of the momentum asymmetry alone.

The author would like to thank Wojtek Broniowski and Wojtek Florkowski for
useful discussions.

\end{document}